\documentclass[aps,prl,twocolumn,amsmath,amssymb,footinbib,showpacs,longbibliography,superscriptaddress]{revtex4-1}

\usepackage[english]{babel}
\usepackage{latexsym}
\usepackage{graphics}
\usepackage{subfigure}
\usepackage{epsfig}
\usepackage{color}
\usepackage{hyperref}
\usepackage{braket} 
\usepackage[T1]{fontenc}
\usepackage[latin9]{inputenc}
\usepackage{amstext}
\usepackage{amssymb}
\usepackage{graphicx}
\usepackage{esint}
\usepackage{braket}
\usepackage{babel}
\usepackage{amsmath}
\usepackage{siunitx}
\usepackage{ulem}

\newcommand{\ie}{i.e.}

\newcommand{\Er}{E_{\textrm{r}}}

\newcommand{\kB}{k_{\textrm{\scriptsize {B}}}}

\newcommand{\ns}{n_{\textrm{s}}}

\newcommand{\rr}{\mathbf{r}}

\newcommand{\asc}{a_{\textrm{\tiny 3D}}}
\newcommand{\aho}{l_\perp}

\newcommand{\gTwoD}{g_{\textrm{\tiny 2D}}}
\newcommand{\gOneD}{g_{\textrm{\tiny 1D}}}

\newcommand{\fc}{f_\textrm{c}}
\newcommand{\Tc}{T_\textrm{c}}
\newcommand{\lamT}{\lambda_\textrm{T}}

\newcommand{\Jscience}{Science}

\newcommand{\Jprl}{Phys. Rev. Lett.}

\newcommand{\Jpre}{Phys. Rev. E}
\newcommand{\Jrmp}{Rev. Mod. Phys.}

\newcommand{\JjphysC}{J. Phys. C: Solid State Phys.}

\begin{document}

\title[2D-1D Dimensional Crossover in Ultracold Bosons]{Experimental Observation of the 2D-1D Dimensional Crossover in Strongly Interacting Ultracold Bosons}


\author{ Yanliang  Guo }
\thanks{These authors contributed equally to this work.}

\affiliation{Institut f{\"u}r Experimentalphysik und Zentrum f{\"u}r Quantenphysik, Universit{\"a}t Innsbruck, Technikerstra{\ss}e 25, Innsbruck, 6020, Austria}

\author{ Hepeng  Yao }
\thanks{These authors contributed equally to this work.}

\affiliation{DQMP, University of Geneva, 24 Quai Ernest-Ansermet, Geneva, CH-1211 ,  Switzerland}

\author{ Satwik  Ramanjanappa}

\affiliation{Institut f{\"u}r Experimentalphysik und Zentrum f{\"u}r Quantenphysik, Universit{\"a}t Innsbruck, Technikerstra{\ss}e 25, Innsbruck, 6020, Austria}

\author{ Sudipta  Dhar}

\affiliation{Institut f{\"u}r Experimentalphysik und Zentrum f{\"u}r Quantenphysik, Universit{\"a}t Innsbruck, Technikerstra{\ss}e 25, Innsbruck, 6020, Austria}

\author{ Milena  Horvath}

\affiliation{Institut f{\"u}r Experimentalphysik und Zentrum f{\"u}r Quantenphysik, Universit{\"a}t Innsbruck, Technikerstra{\ss}e 25, Innsbruck, 6020, Austria}

\author{ Lorenzo  Pizzino}
\affiliation{DQMP, University of Geneva, 24 Quai Ernest-Ansermet, Geneva, CH-1211 ,  Switzerland}

\author{ Thierry Giamarchi}
\affiliation{DQMP, University of Geneva, 24 Quai Ernest-Ansermet, Geneva, CH-1211 ,  Switzerland}

\author{ Manuele  Landini}

\affiliation{Institut f{\"u}r Experimentalphysik und Zentrum f{\"u}r Quantenphysik, Universit{\"a}t Innsbruck, Technikerstra{\ss}e 25, Innsbruck, 6020, Austria}

\author{ Hanns-Christoph  N{\"a}gerl}\email{christoph.naegerl@uibk.ac.at}


\begin{abstract}
Dimensionality plays an essential role in determining the nature and properties of a physical system. For quantum systems the impact of interactions and fluctuations is enhanced in lower dimensions, leading to a great diversity of genuine quantum effects for reduced dimensionality. In most cases, the dimension is fixed to some integer value. Here, we experimentally probe the dimensional crossover from two to one dimension using strongly interacting ultracold bosons in variable lattice potentials and compare the data to ab-initio theory that takes into account non-homogeneous trapping and non-zero temperature. From a precise measurement of the momentum distribution we analyze the characteristic decay of the one-body correlation function in the two dimensionalities and then track how the decay is modified in the crossover. A varying two-slope structure is revealed, reflecting the fact that the particles see their dimensionality as being one or two depending on whether they are probed on short or long distances, respectively. Our observations demonstrate how quantum properties in the strongly-correlated regime evolve in the dimensional crossover as a result of the interplay between dimensionality, interactions, and temperature.
\end{abstract}

\keywords{Dimensional crossover, ultracold atoms, superfluids, correlation functions}

\maketitle
\def\thefootnote{*}\footnotetext{These authors contributed equally to this work}\def\thefootnote{\arabic{footnote}}
Understanding the effects of interactions in a quantum system is one of the central problems of the field of quantum many-body physics. Interactions lead to phases of matter with remarkable properties, such as superfluidity or fractional quantum Hall conductance \cite{tsui_FQHE,laughlin_liquid}. Quite crucially such effects are intimately linked to the dimensionality of the system. Lower dimensions usually 
reinforce the effects of interactions, as well as of quantum and thermal fluctuations, allowing for even greater deviations from the physics of non-interacting systems. Taking interacting bosons as a comparatively simple example, these go from robust three-dimensional (3D) superfluids with Bogoliubov quasi-particle excitations \cite{pitaevskii_becbook} to quasi-long-range ordering with Berezinski-Kosterlitz-Thouless (BKT) topological phase transitions in 2D \cite{kosterlitz1973,hadzibabic_ENS_experiment_BKT} and to Tomonaga-Luttinger liquids in 1D \cite{giamarchi_book_1d}, where quasi-particles do not exist and collective fermionization occurs \cite{girardeau_tonks_gas,paredes_tonks_experiment,kinoshita2004} for strong interactions.

In most systems, the dimensionality is fixed once for all and the various phenomena are studied independently. There is, however, a very interesting class of systems for which the effective dimensionality can be controlled via  parameters such as temperature, confinement lengthscale, or observational lengthscale. Examples of systems that feature such a dimensional crossover \cite{giamarchi_qpt_carr}
are organic conductors and superconductors \cite{lebed_book_1d,giamarchi_review_chemrev} made of weakly coupled one-dimensional chains, or high-temperature superconductors made of weakly coupled planes for fermions \cite{orenstein_review_hightc}, or  coupled chains and ladders for bosons \cite{ho04_deconfined_bec,cazalilla-coupled1D-2006,klanjsek_bpcp,hong_DIMPY,bollmark-crossoverD-2020}. Needless to say, in such systems one of the crucial questions is 
to understand how the dimensionally different phases are linked together upon a variation of temperature, tunnel coupling, or distance, and how the high-dimensional phases are influenced by the exotic low-dimensional physics coming from the high-temperature or short-distance regime. Doing so requires however an excellent degree of control and the right range of parameters such as tunnelling vs temperature, which is a strong challenge in material research for condensed-matter realizations. 

In this work, guided by a recent theoretical study~\cite{yao-crossoverD-2022}, we present the first experimental observation of the quantum correlation properties for strongly-interacting cold atomic bosons in the dimensional crossover from 2D to 1D. In recent years, cold atoms \cite{pitaevskii_becbook} have provided, due to their remarkable degree of control of the microscopic Hamiltonians, a remarkable alternative playground to address strongly-correlated quantum matter \cite{bloch-review-2008}. First studies have been carried out for weakly interacting gases addressing non-equilibrium dynamics~\cite{jorg-crossD-2007,li-1D-T-2020,Jorg-hydrodynamics-crossD-2021}, excitation spectra~\cite{stoferle-crossoverD-2004} and transverse superfluidity~\cite{dalibard-crossoverD-2023} in regimes when the dimension is not strictly integer.  
However, testing the quantum correlation properties predicted in Ref.~\cite{yao-crossoverD-2022} is not possible when interactions are weak. Our Cesium quantum gas, on the contrary, offers the possibility of strong correlations and thus to have a clear distinction between the two integer dimensionalities, which prompts us to follow the evolution during the dimensional crossover. With quantum Monte Carlo (QMC) calculations, we benchmark the experimental data at both $D=2$ and $1$. Within the dimensional crossover, we observe a two-regime structure for the first-order correlation function. We explicitly show how the system evolves from a 2D gas with BKT algebraic decay for the correlation function to a system of coupled 1D tubes with exponential decay, and how it finally becomes a system of incoherently coupled 1D tubes.

\begin{figure*}[t!] 
\centering
\includegraphics[width=1.7\columnwidth]{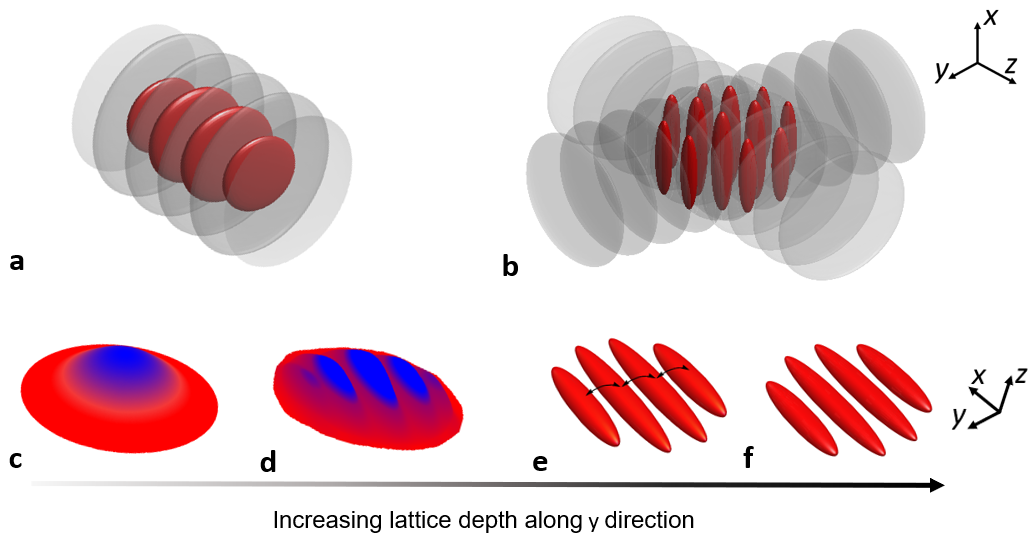}
\caption{\label{fig:sketch} 
\textbf{Conceptual sketch of the experiment.} Starting from a 3D BEC, we generate an ensemble of low-dimensional units, namely a. 2D layers and b. 1D tubes. (c-f) Schematic plots for the evolution of one particular layer during the dimensional crossover, where the quantum gas goes from 2D (c), to modulated 2D (d), to coherently-coupled 1D (e), and then to incoherently-coupled 1D (f) by continuously increasing the lattice depth. The blue color in (c-d) indicates the high-density regime where superfluid regions appear.}
\end{figure*}

The experimental sequence starts with an interaction-tunable 3D Bose-Einstein condensate (BEC) containing typically $1.5\!\times\!10^5$ Cs atoms prepared in the lowest magnetic hyperfine state $\vert F,m_F \rangle=\vert 3,3 \rangle$, held in a crossed-beam dipole trap and levitated against gravity by a magnetic field gradient ~\cite{Kraemer2004}. The temperature of the 3D BEC is \SI{10(1)}{\nano\kelvin} as determined from a small non-condensed fraction in time-of-flight (TOF) measurements. With the 3D s-wave scattering length set to $a_{\textrm{\tiny 3D}}\approx200 a_0$, the BEC is in the Thomas-Fermi regime. A 2D optical lattice with lattice beams along the $y$ and $z$ directions with a lattice spacing $a\!=\!\lambda/2\!\!=\!\!532$~nm is gradually ramped up in \SI{500}{\milli\second} to a target value, with $\lambda$ the wavelength of the lattice light. The lattice depth along the $z$ direction is always set to $V_z=30E_r$, with $\Er \!=\! \pi^2\hbar^2/(2ma^2)$ the recoil energy, resulting in tight transversal harmonic trapping with trap frequency $\omega_{\rm z}/2\pi\!=\!\SI{11}{\kilo\hertz}$. Along the $y$-direction, the lattice depth $V_y$ is varied in the range from $0$ to $30\Er$. At the two limiting values $V_y/\Er\!=\!0$ and $30$ the system forms an ensemble of 2D layers and an array of 1D tubes, respectively, see Fig.~\ref{fig:sketch}. With the lattice beams providing also some transversal confinement, the trapping frequency $\omega_{x}$ along the third direction varies from $\omega_{x}/2\pi \!=\!\SI{10.1(2)}{\hertz}$ for $V_y\!=\!0$ to $\omega_{x}/2\pi\!=\!\SI{14.3(2)}{\hertz}$ for $V_y\!=\!30\Er$.

After loading the atoms into the lattice, the offset magnetic field $B$ is ramped adiabatically to tune $a_{\textrm{\tiny 3D}}$ to $620 a_0$. This sets the 2D interaction parameter to $\gamma_{\textrm{\tiny 2D}}\!\!=\!\!1.5$ in the strictly-2D regime and the Lieb-Linger parameter to $\gamma_{\textrm{\tiny 1D}}\!=\!40$ in the strictly-1D regime (see Methods), putting our systems in the strongly-interacting regime in both integer dimensions~\cite{bloch-review-2008,hadzibabic-2Dgas-2011,cazalilla-1dreview-2011}. In 1D, the interaction is strong enough that the so-called Tonks-Girardeau regime is entered, where the 1D bosons show fermionized behavior~\cite{girardeau_tonks_gas,paredes_tonks_experiment,kinoshita2004}. Given the lattice along the $y$ direction, we are able to accurately control the dimensionality of the strongly-interacting bosonic system. By increasing $V_y$ continuously, the system goes through the whole dimensional crossover from 2D to 1D. As $V_y$ rises, each single layer goes through various regimes, namely: strictly 2D (S2D), modulated 2D (M2D), coupled 1D tubes (C1D) and isolated 1D tubes (I1D), as illustrated in Fig.~\ref{fig:sketch} (c-f).

The main quantity we concentrate on is the one-body correlation function $g^{(1)}(x,x',y,y')\!=\!\langle \hat{\Psi}^\dagger(x', y')\hat{\Psi}(x, y)\rangle$, which allows us to extract the quantum coherence properties of the cold atomic gas. In our experiment, $g^{(1)}$ is measured via the TOF technique. After loading the lattice, we hold the gas in the trap for \SI{50}{\milli\second}. We remove all trapping potentials for 56-ms TOF with zeroed interactions \cite{weber2003}, followed by absorption imaging. The TOF duration is long enough to satisfy the far-field approximation. Given the interaction zeroing, the measured density distribution is a faithful representation of the 2D resp. 1D momentum distribution $n(k_x,k_y)$. Fourier transform then gives the information of $g^{(1)}$. Specifically, we obtain the integrated correlation function
\begin{equation}\label{eq:g1-integrated}
G^{(1)}(x,y)=\iint dx^{\prime}dy^{\prime} \langle \Psi^\dagger(x^{\prime}+x,y^{\prime}+y)\Psi(x^{\prime},y^{\prime})\rangle.
\end{equation}

\begin{figure*}[t!]
\centering
\includegraphics[width = 1.8\columnwidth]{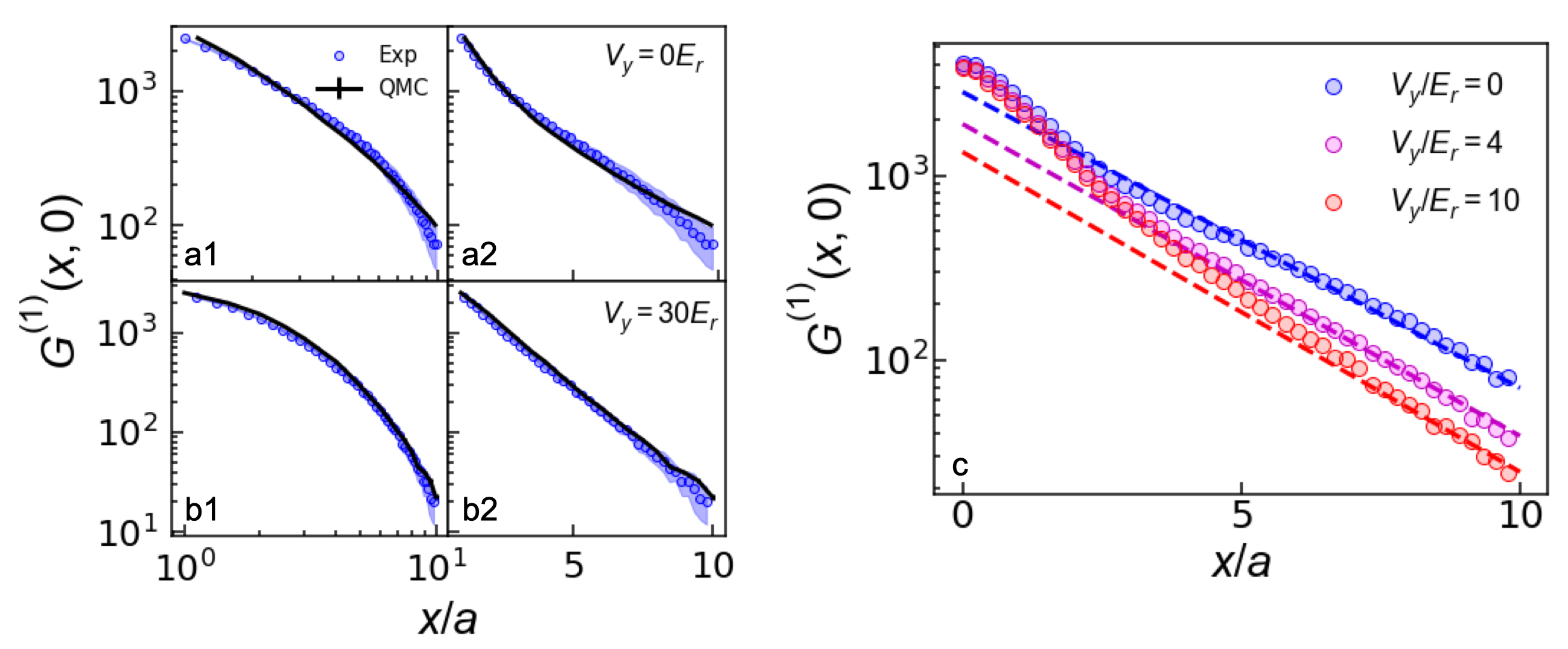}
\caption{\label{fig2-g1}
\textbf{Characteristic decay of the longitudinal correlation function $G^{(1)}(x,0)$}. Measured $G^{(1)}(x,0)$ in the two integer dimensions (a1-b2) and in the dimensional crossover (c) as a function of longitudinal distance $x/a$.  In a1 and a2 (b1 and b2), the data (circles) for the 2D (1D) case with $V_y/\Er\!=\!0$ ($V_y/\Er\!=\!30$) is plotted on a log-log resp. semi-log scale, averaged over 30 repetitions of the experiment, and is compared to the QMC calculations (solid lines). For the data in c, the strength of the transversal lattice $V_y$ is varied as indicated. The dashed lines are linear fits of the long-range behavior in the semi-log scale. Their slopes vary by less than 10\%. 
}
\end{figure*}

The typical decay with distance is very distinct and its analytical form in the homogeneous case for the different integer dimensionalities has been calculated. In 2D, $g^{(1)}$ is expected to decay as\begin{equation}\label{eq:g1-2D-alg}
g^{(1)}_{\textrm{\tiny 2D}}(\rr,0)\sim \left\{
\begin{aligned}
\vert \rr \vert^{-\alpha_{\textrm{\tiny 2D}}},\ T<\Tc \\
e^{-\eta_{\textrm{\tiny 2D}} \vert \rr \vert},\ T>\Tc.
\end{aligned}
\right.
\end{equation}
with 2D distance $\rr$. Below the critical temperature $\Tc$ given by $\ns \lamT^2(\Tc) \!=\! 4$ the decay is algebraic with a BKT-type exponent $\alpha_{\textrm{\tiny 2D}}\!=\!1/\ns\lamT^2$, where $\ns$ is the 2D superfluid density and $\lamT$ is the de Broglie wavelength~\cite{bloch-review-2008,hadzibabic-2Dgas-2011}. Above $\Tc$, the decay is exponential with a thermal exponent $\eta_{\textrm{\tiny 2D}}\!=\!\frac{\sqrt{4\pi}}{\lamT e^{n\lamT^2/2}}$ where $n$ is the 2D density.  
In the strict 1D limit, $g^{(1)}$ is expected to exhibit various decay regimes at different distances
\begin{equation}\label{eq:g1-1D-alg}
g^{(1)}_{\textrm{\tiny 1D}}(x,0)\sim \left\{
\begin{aligned}
x^{-\alpha_{\textrm{\tiny 1D}}}&,&\ x_0<x<\xi(T) \\
e^{- \eta_{\textrm{\tiny 1D}} x}&,&\ x>\xi(T)
\end{aligned}
\right.
\end{equation}
Above a short-range cutoff $x_0$ (which is less than the typical experimental  resolution) $ g^{(1)}_{\textrm{\tiny 1D}}$ decays algebraically with a Tomonaga-Luttinger-liquid-type exponent $\alpha_{\textrm{\tiny 1D}}\!=\! 1/2K$, where $K$ is the Luttinger parameter. For larger distances above the thermal length $\xi(T)\!=\!2uK/(\pi \kB T)$, it decays exponentially with exponent $\eta_{\textrm{\tiny 1D}}\!=\!1/\xi(T)$. Here, $u$ is the sound velocity in the 1D gas~\cite{giamarchi_book_1d,cazalilla-1dreview-2011}. Notably, for $T>0$, the regime of the exponential decay always exists.

Our results for $G^{(1)}(x,y\!=\!0)$ of the trapped system in the two integer-dimension limits and in the crossover regime are displayed in Fig.~\ref{fig2-g1}. In 2D, the decay is algebraic for short distances and then turns over into an exponential decay. In 1D, the decay is clearly exponential. The QMC calculations recover these two regimes well. In the dimensional crossover regime, we observe a two-structure decay pattern as shown in Fig.~\ref{fig2-g1} (c) and discussed in more detail below. 

For the 2D case, our data suggests that the change of the decay behavior from an algebraic to an exponential happens around $r\sim 5a$. Fitting the data with $G^{(1)}(x,0) \sim x^{-\alpha}$ in the range of $r<5a$ and $G^{(1)}(x,0) \sim e^{-\eta_{\textrm{\tiny 2D}} x/a}$ in the range of $5a<r<10a$ gives $\alpha\!=\!1.1(1)$ and $\eta_{\textrm{\tiny 2D}}\!=\!0.36(1)$. The BKT-type algebraic decay is consistent with the existence of a superfluid core in the center of the trap (blue region in Fig.~\ref{fig:sketch} (c)). At larger radii, a ring of normal fluid is formed, which contributes to the exponential decay (red region in Fig.~\ref{fig:sketch} (c)).

The validity of such a simple model approach can be checked by QMC calculations. We use the continuous space path-integral MC method to simulate the system within the grand-canonical ensemble at finite temperature. In each layer, the system is modeled by the elementary many-body Hamiltonian
\begin{equation}\label{eq:Hamiltonian}
 \hat{H} \!=\! \sum_j \left[ -\frac{\hbar^2}{2m} \nabla^2_j + V(\hat{\rr}_j) \right] + \sum_{j<k} U(\hat{\rr}_j - \hat{\rr}_k)
 \end{equation}
 with $\hat{\rr}_j\!=\!(x_j,y_j)$ the position of the $j$-th particle. The term $U(\hat{\rr})$ models the short-range repulsive two-body interaction, see Methods. The external potential reads $V(\rr) \!=\! m \omega^2 \vert \rr \vert^2/2 + V_y \cos^2 (\pi y/a) $, consisting of a 2D harmonic trap and a unidirectional lattice. The correlation function $G^{(1)}(x,y)$ is computed by the worm algorithm implementation ~\cite{gautier-2Dquasicrystal-2021,yao-crossoverD-2022}. We simulate a single layer given the experimental conditions, namely with a weighted atom number $N_w\!=\!4000$ (see Methods and Refs.~\cite{Haller2011,meinert-1Dexcitation-2015}) and for strong interactions $a_{\textrm{\tiny 3D}}\!\!=\!\!620 a_0$. We find that our experimental data is well fitted by the QMC calculation when the temperature is set to $T\!\!=\!\!\SI{16}{\nano\kelvin}$.

In the 1D limit, the experimental data shows an exponential decay over the full range $0<x<10a$, see Fig.~\ref{fig2-g1} b1 and b2. Given the strong interactions, the 1D system is deeply in the Tonks-Girardeau regime. In this regime, the exponential decay of the correlation function has been calculated under trapping conditions~\cite{minguzzi2022-1D-slope}, giving $G^{(1)}(x)\sim e^{-\eta_{\textrm{\tiny 1D}} x/a}$, with $\eta_{\textrm{\tiny 1D}}\!=\!m\kB Ta/2\hbar^2 n_0$ and $n_0$ the 1D particle density in the center of the trap. Fitting the experimental data with this formula gives $\eta_{\textrm{\tiny 1D}}\!=\!0.53(1)$. This is consistent with the QMC simulation with weighted atom number $N_{w}\!=\!25$ and center particle density $n_0 a\!=\!0.51$ for a temperature of $T\!\!=\!\!\SI{7}{\nano\kelvin}$. As a cross check, by using this temperature value and $n_0 a\!=\!0.51$ for the central density, the analytical formula gives $\eta_{\textrm{estimate}}\!=\! 0.52$. This fits well with the experimental data. Note that, thanks to the strong interactions, the 1D decay constant differs greatly from the one in 2D. Note also that the temperatures in the three integer dimensions, $10(1)$, $16(1)$, and $7(1)$ \SI{}{\nano\kelvin}, determined as discussed above, are strikingly different. This will be the subject of a forthcoming publication \cite{Entropy}. 

\begin{figure}[t!]
\centering
\includegraphics[width = 0.88\columnwidth]{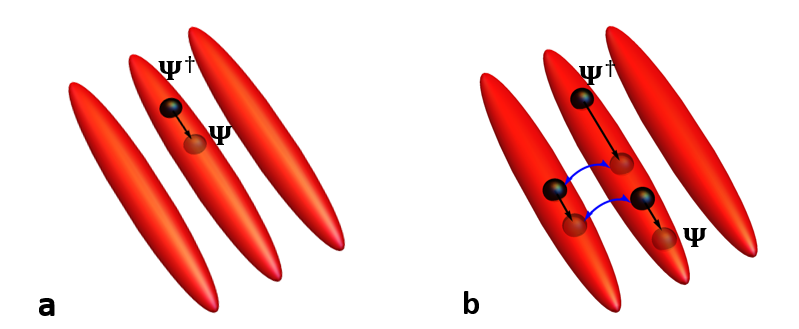}
\caption{\label{fig3}
\textbf{Illustration of the mechanism behind the first-order correlation function and its interpretation.} (a) On short distances, the particles have less chance to tunnel into the neighboring tubes. They perceive their world as 1D. (b) For longer distances, the particles have more opportunity to tunnel between the tubes. The 2D character takes over.
}
\end{figure}

Our data shows that both the short-range and the long-range behavior of $G^{(1)}(x,0)$ is subject to a significant modification as the dimensionality is switched from 2D to 1D. But how does this behavior evolve as the dimensionality is continuously changed, i.e., in the dimensional crossover regime? One naive guess could be the correlations decay faster at \textbf{all} distances as the full pattern smoothly evolves from 2D to 1D. Remarkably, we observe a different behavior, namely that the short-range behavior is governed by a stronger decay, while the long-range behavior still remains the same as in strictly-2D. This is seen in our data in Fig.~\ref{fig2-g1} c as the transversal potential $V_y$ is set to intermediate values. Although the long-range 2D decay region shrinks as the potential $V_y$ is increased, it maintains the same decay $G^{(1)} (x,0) \sim e^{-\eta_{\textrm{\tiny 2D}} x}$ in 2D.

This behavior can be explained by the sketch in Fig.~\ref{fig3}, similarly to the statement made for homogeneous systems in Ref.~\cite{yao-crossoverD-2022}, and it is a direct visualisation of the mechanism that is behind the correlation function $g^{(1)}(x,x',y,y)\!=\!\langle \hat{\Psi}^\dagger(x', y)\hat{\Psi}(x, y)\rangle$: The probability amplitude of creating a particle at position $x'$ and then destroying it at position $x$. When $\lvert x-x' \rvert$ is small, the particle will have less chance to tunnel into one of the neighboring tubes, and less so for higher values of the lattice depth. Thus, for short distances, the particles see themselves as living in 1D. Conversely, when $\lvert x-x' \rvert$ is large, the particles have more opportunity to tunnel between the tubes. In essence, they perceive their world as being 2D. As a consequence, the short-range character of the system will mimic the 1D behavior first, while the long-range behavior remains well described by the behaviour observed in 2D, as the transversal confinement is strengthened. 

\begin{figure}[t!] 
\centering
\includegraphics[width=0.8\columnwidth]{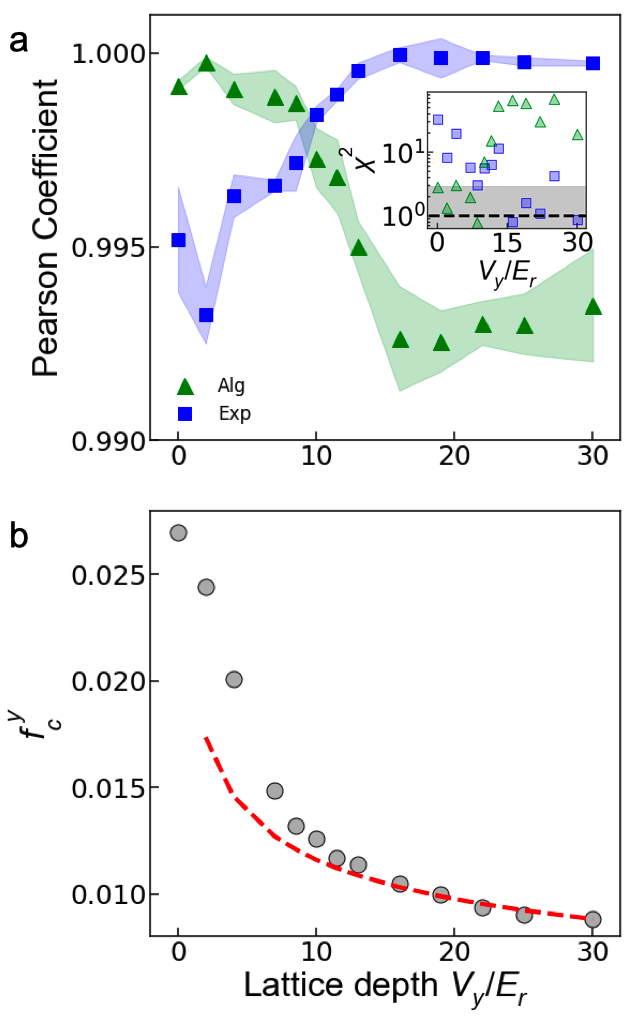}
\caption{\label{fig3-P} 
\textbf{Dimensional crossover analysis.} (a) Pearson correlation coefficient $P$ as a function of the lattice depth $V_y$ for the data in log-log (green triangles) and semi-log scale (blue squares). The shaded regions reflect the error and they are determined by shifting the window to calculate $P$ by $\Delta x\!=\!\pm 0.25a$. The inset gives the $\chi^2$ value for the data, with the grey region indicating the targeted area in view of statistical fluctuations. (b) Transverse zero-momentum fraction $f_c^y$ as a function of $V_y$. The data (circles) is the average of 30 repetitions, and the statistical error bar is smaller than the symbol size. It is compared to the value for the ground state of quantum harmonic oscillator (red dashed line).
}
\end{figure}

We now turn to the short-range behavior of $G^{(1)}(x,0)$ in the dimensional crossover regime. To capture the transition from a 2D-power-law to a 1D-exponential decay, we use a Pearson correlation coefficient analysis by computing $P(X,Y)=\mathrm{cov} (X,Y)/ \sigma_X \sigma_Y$ for the data sets $(x,\log\ G^{(1)})$ and $(\log x,\log\ G^{(1)})$ in the range $1.7<x/a<4.5$. The value $P(X,Y)$ is a measure for the linearity of the dataset, and with the semi-log and log-log rescaled data it allows us to distinguish between algebraic and exponential decay. The results are shown in Fig.~\ref{fig3-P} a. The decay is closer to algebraic when $V_y<10\Er$. Around $V_y=10\Er$, the decay becomes an exponential. Our data can be further examined on the basis of its statistical fluctuations by computing the normalized $\chi^2$ parameter for the algebraic and exponential fits~\cite{murthy-chi2-2015}, see inset of Fig.~\ref{fig3-P}. We compare the data to $1\pm 5\sigma$ (grey area), where $\sigma$ is the standard deviation of $\chi^2$~\cite{bevington-book-1993}. For the two limits $V_y/\Er<7$ and $V_y/\Er>15$, we find that the $\chi^2$ values of the algebraic and exponential fits drop well into the grey area. This agrees with the result of the Pearson analysis.

Our data shows that the crossover from M2D to C1D happens around $10\Er$. This value sets the scale of the tight-binding regime and fits with the theoretical prediction \cite{bloch-review-2008,yao-crossoverD-2022}. This is consistent with the behavior of $\chi^2$ in the range $7<V_y/\Er<15$, where none of the two $\chi^2$ values is close to 1 within reasonable statistical fluctuations. This can be understood in the context of our model, predicting a coexistence of the two decay behaviors. Note that the exact crossing point of the two P values may change for varying temperatures, interaction strengths and particle numbers. 

We finally turn to another quantity that also sheds light onto the dimensional crossover. The transverse zero-momentum fraction $\fc^y\!=\! \sum_{k_x} n(k_x,0)/ \sum_{k_x,k_y} n(k_x,k_y)$ reflects the coherence along the transversal direction~\cite{Plisson2011}, see Methods. As can be seen in Fig.~\ref{fig3-P} b its value strongly drops across the M2D regime. The decay slows down in the C1D regime. Finally, in the I1D regime, it saturates and fits well with the prediction of the zero-momentum component for the ground state of the quantum harmonic oscillator. Such a prediction fails in the M2D regime because the system evolves into a higher dimensionality and gains more coherence. The behavior of our data is in agreement with the prediction for the homogeneous case~\cite{yao-crossoverD-2022}.

In conclusion, we have presented an experimental characterization of the 2D-1D dimensional crossover for strongly-interacting bosons from the BKT to the Tonks-Girardeau regime. We have tracked the continuous evolution of the correlation function and the transverse zero-momentum fraction and we have shown experimentally that the short-range behavior acquires the 1D character first, while the long-range behaviour remains 2D, in agreement with the theoretical expectation. The short-range decay of $g^{(1)}$ evolves continuously from algebraic to exponential as captured by the Pearson correlation coefficient. Our experimental measurements provide the first evidence of such a two-region structure during the dimensional crossover. In principle, such behavior is also expected at the dimensional crossover between other dimensionalities (\ie, 3D-1D, 3D-2D), which could lead to further extensions of our observations. We expect that our work will trigger the study of the dimensional crossover in other types of systems, both on the bosonic and the fermionic side. Understanding such a crossover for fermionic systems will in particular be directly relevant for a host of condensed-matter realizations, such as quantum spin chains or organic and high-temperature superconductors, which exhibit the very anisotropic structure discussed in this paper, as well as very unusual properties.

\acknowledgments
The Innsbruck team acknowledges funding by a Wittgenstein prize grant under project number Z336-N36 and by the European Research Council (ERC) under project number 789017. This research was funded in part by the Austrian Science Fund (FWF) W1259-N27 and MH thanks the doctoral school ALM for hospitality. This work is also supported by the Swiss National Science Foundation under grant number 200020-188687. Numerical calculations make use of the ALPS scheduler library and statistical analysis tools~\cite{troyer1998,ALPS2007,ALPS2011}.

\textbf{Data access statement}

The data that support the findings of this study are made publicly available from Zenodo by the authors at \cite{Zenodo}.

\bigskip

\section*{Methods}

\subsection*{Preparation of the experiment }
\label{Preparation of the experiment}

Our experiment starts with a 3D Bose-Einstein condensate (BEC) containing typically $1.5\times10^5$ Cs atoms prepared in the lowest magnetic hyperfine state $\vert F,m_F \rangle=\vert 3,3 \rangle$, held in a crossed-beam dipole trap with a 3D trapping frequency $\omega_{\rm x,y,z}/2\pi=(15,8,17)\SI{}{\hertz}$. It is levitated against gravity by means of a magnetic field gradient $\nabla B\sim 31.1$ G/cm oriented along the vertical direction, \ie, x-direction in the main text. The 3D s-wave scattering length set to $a_{\textrm{\tiny 3D}}\approx200 a_0$ where the BEC is in the Thomas-Fermi regime. The details of the trapping and cooling procedures are described in ~\cite{weber2003,Kraemer2004}. To prepare an ensemble of 2D Bose gases or an array of 1D gases, we adiabatically load the BEC into an optical lattice, generated from two orthogonally and horizontally propagating retro-reflected laser beams at a wavelength $\lambda = 1064.5$ nm. After the loading procedure the lattice depth in z-direction is always set to $V_z=30E_r$, with $\Er \!=\! \pi^2\hbar^2/(2ma^2)= h\times 1.325$ kHz denoting the recoil energy for Cs atoms. The lattice results in a tight transversal harmonic trapping frequency of $\omega_{z}/2\pi\!=\!\SI{11}{\kilo\hertz}$. The weak radial confinement in 2D and the longitudinal one in 1D caused by the combined lattice trapping potentials give $\omega_{x}^{\textrm{\tiny 2D}}/2\pi\!=\!\SI{10.1(2)}{\hertz}$ and $\omega_{x}^{\textrm{\tiny 1D}}/2\pi\!=\!\SI{14.3(2)}{\hertz}$. During the loading process almost all layers (2D) or tubes (1D) are in the Thomas-Fermi (TF) regime for weak repulsive interactions, so we can calculate the initial occupation number in each layer in 2D or each tube in 1D case through the global chemical potential and the total atom number~\cite{Haller2011,meinert-1Dexcitation-2015}. With the calculation $\bar{N}= \sum_{i} N_{i}^2/\sum_{i} N_{i}$ and $N_i$ the atom number of the $i$-th tube (or layer), we obtain the weighted average number $\overline{N}_{\textrm{\tiny 2D}}=4000$ in 2D and  $\overline{N}_{\textrm{\tiny 1D}}=25$ in 1D.

In the lattice, we adiabatically raise the 3D scattering length to $a_{\textrm{\tiny 3D}}\!=\!620a_0$ by means of a broad magnetic Feshbach resonance \cite{weber2003}. The ramp time ($50$ ms) is chosen carefully, i.e., slow enough to avoid any excitations of breathing modes in the gas. In both the 2D and 1D regimes, we calculate the coupling constants $g$ to
\begin{align}\label{eq:coupling-petrov}
\gTwoD \simeq \frac{2\hbar^2 \sqrt{2\pi}}{m \aho/\asc+{1/\sqrt{2\pi} \ln (1/\pi q^2 \aho^2)}}, \\ 
\gOneD= \frac{2 \hbar^2 \asc}{m\aho^2} \bigg(1-\frac{1.036\asc}{\aho}\bigg)^{-1}
\end{align}
with $\aho= \sqrt{\hbar/m\omega_\perp}$ the characteristic transverse length, $q=\sqrt{2m\vert\mu\vert/\hbar^2}$ the quasi-momentum, and $\mu$ the chemical potential.
The interaction regime of the system can further be captured by the 2D interaction parameter $\gamma_{\textrm{\tiny 2D}}=m \gTwoD /\hbar^2=1.5$ and 1D Lieb-Liniger parameter $\gamma_{\textrm{\tiny 1D}}=m \gOneD /\hbar^2na=40$. The criteria for the strongly interacting regimes in 2D and 1D are given by $\gamma_{\textrm{\tiny 2D}} \geq 1$ and $\gamma_{\textrm{\tiny 1D}}\gg 1$~~\cite{bloch-review-2008,hadzibabic-2Dgas-2011,cazalilla-1dreview-2011}, respectively.

\subsection*{  Amplitude-amplitude correlation function $G^{(1)}(x,0)$}
\label{app:Fourier transform}

In the present work, we study the integrated correlation function of the trapped system
\begin{equation}
\label{eq:G1-def}
G^{(1)}(x,y)=\iint dx^{\prime}dy^{\prime}\langle \Psi^\dagger(x^{\prime}+x,y^{\prime}+y)\Psi(x^{\prime},y^{\prime})\rangle.
\end{equation}
For a homogeneous system, this value can be linked with the one-body correlation function $g^{(1)}(x,y)=\langle \Psi^\dagger(x,y)\Psi(0,0)\rangle$ by $G^{(1)}(x,y)=g^{(1)}(x,y)L_xL_y$, with $L_x,L_y$ the system sizes along the $x,y$-directions. Note that, due to the break of translational invariance in the trapped system, such a relation is not valid anymore and we have to focus on the quantity $G^{(1)}(x,y)$. Thanks to the correspondence of reciprocal space, it can be linked to the momentum distribution $n(k)$ by Fourier transform
\begin{equation}
\label{eq:G1-nk}
G^{(1)}(x,y)=\sum_{k_x,k_y} n(k_x,k_y) e^{-i(k_xx+k_yy)}\Delta k_x\Delta k_y.
\end{equation}
In the experiment, the momentum distribution $n(k_x,k_y)$ is obtained by performing a time-of-flight measurement while $\Delta k_x$ and $\Delta k_y$ are the resolutions in momentum space, limited by the imaging camera's pixel size. For this, applying a discrete Fourier transform on $n(k_x)\Delta k_x$ results in the integrated amplitude-amplitude correlation function $G^{(1)}(x,0)$ as plotted in Fig~\ref{fig2-g1}.


\subsection*{  Resolving the image angle for the zero-momentum fraction}

\begin{figure}[h] 
\includegraphics[width=0.95\columnwidth]{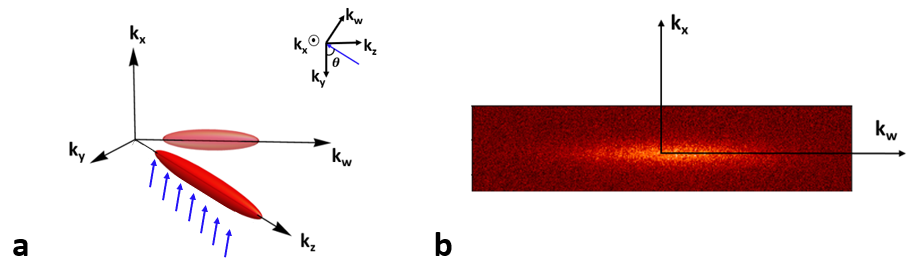}
\caption{\label{fig:sketch111}
\textbf{A schematic of our imaging setup and an example of the image.} (a) The vectors $k_y$, $k_z$, $k_w$ and the imaging direction (blue arrows) all lie in one plane, with $k_x$ perpendicular to this plane, see also inset. The red 3D ellipsoid along the $k_z$ direction indicates the atomic cloud after TOF starting from an ensemble of 2D layers for $V_{y} = 0 E_{r}$. The  light red 2D ellipsoid along $k_w$ direction is the shadow in our absorption image. (b) An example of a projected image after TOF for $V_{y} = 0 E_{r}$. }

\end{figure}

Our setup consists of two lattice beams perpendicular to each other along the $y$ and $z$ axes. There is an angle of $\theta \sim 57^\circ$  between the propagation axis of the imaging beam and the $y$-direction, see Fig. \ref{fig:sketch111}.  To determine the momentum distribution $n_{\textrm{\tiny y}}(k_y)$, we need to deconvolve the measured one $n_{\textrm{\tiny w}}(k_w)$. Specifically, the vectors $k_w$, $k_z$ and $k_{y}$ lie in one plane and hence $k_{y}$ can be expressed in terms of $k_{w}$ and $k_{z}$. The measured $n_{\textrm{\tiny w}}(k_w)$ can thus be written as a product of the momentum distributions $n_{\textrm{\tiny y}}(k_y)$ along $y$- and $n_{\textrm{\tiny z}}(k_z)$ along $z$-direction
\begin{equation}
    n_{\textrm{\tiny w}}(k_{w}) = \int  n_{\textrm{\tiny z}}(k_{z})n_{\textrm{\tiny y}}(\frac{k_{w}}{\sin \theta}-\frac{k_{z}}{\tan \theta}) dk_{z},
    \label{convoeqn2}
\end{equation}

which is a convolution of $n_{\textrm{\tiny y}}$ and $n_{\textrm{\tiny z}}$. Since $V_z=30E_r$, $n_{\textrm{\tiny z}}(k_z)$ is well approximated by the ground state of the harmonic oscillator. Thus, deconvolving and applying the inverse Fourier transform on the functions $n_{\textrm{\tiny z}}$ and $n_{\textrm{\tiny w}}$ gives us $n_{\textrm{\tiny y}}$. We then take the atom number within $\Delta k_y=2\pi/L_y$ to calculate the zero-momentum fraction.

\subsection*{The quantum Monte Carlo calculation}

The numerical computation of the one-body correlation function $G^{(1)}(x,y)$ is carried out by the path integral Monte Carlo method~\cite{ceperley-PIMC-1995}. Similarly as Refs.~\cite{yao-tancontact-2018,yao-boseglass-2020,yao-crossoverD-2022}, we simulate the system with the Hamiltonian in Eq. \ref{eq:Hamiltonian} in the main text within the grand canonical ensemble, at finite temperature $T$, interaction strength $\gTwoD$ and chemical potential $\mu$ (equivalently particle number $N$). Notably, the two-body interaction propagator is generated by the pair-product approximation and generalized to any interaction regimes, for details see Ref.~\cite{gautier-2Dquasicrystal-2021}. We take into account the presence of both the continuous optical lattice and the harmonic trap, with the same parameters as for the experimental setup. By properly adapting the numerical parameters, we can simulate 2D systems up to sizes of $130a$ and particle numbers $5000$. Moreover, thanks to the worm algorithm implementations~\cite{boninsegni-worm-short-2006,boninsegni-worm-long-2006}, we can extended our simulations to the open wordline G sectors, where the statistics of creation and annihilation operators with open ends at $(x',y')$ and $(x'+x,y'+y)$ are counted. This enables us to calculate the one-body correlation function defined in Eq.\ref{eq:g1-integrated} of the main text. For symmetry reasons, it is sufficient to calculate it to half of the system size. The numerical calculations make use of the ALPS scheduler library and statistical analysis tools~\cite{troyer1998,ALPS2007,ALPS2011}.


\begin{thebibliography}{10}
\bibitem{Zenodo}
{Data set is available from {Z}enodo at doi:10.5281/zenodo.8167239}.
\end{thebibliography}

\end{document}